\documentclass[letter]{aa}  
\bibpunct{(}{)}{;}{a}{}{,} 
\usepackage{graphicx}
\usepackage{txfonts}
\usepackage{natbib}
\newcommand{\be}{\begin{equation}}
\newcommand{\ee}{\end{equation}}
\newcommand{\bea}{\begin{eqnarray}}
\newcommand{\eea}{\end{eqnarray}}

\newcommand{\ha}{H$\alpha$}

\begin{document} 

\authorrunning{C. E. Alissandrakis etal}
\title{A first look at the submillimeter Sun with ALMA
}
\author{C. E. Alissandrakis$^1$, T. S. Bastian$^2$, A. Nindos$^1$
}

\institute{Department of Physics, University of Ioannina, GR-45110 Ioannina, 
Greece\\
\email{calissan@uoi.gr}
\and
National Radio Astronomy Observatory (NRAO), 520 Edgemont Road, Charlottesville, VA 22903, USA
}

\date{Received ...; accepted ...}

  \abstract
{
We present the first full-disk solar images obtained with the Atacama Large Millimeter/submillimeter Array (ALMA) in Band 7 (0.86 mm; 347 GHz). In spite of the low spatial resolution (21\arcsec), several interesting results were obtained. During our observation, the sun was practically devoid of active regions. Quiet Sun structures on the disk are similar to those in Atmospheric Imaging Assembly (AIA) images at 1600\,\AA\ and 304\,\AA, after the latter are smoothed to the ALMA resolution, as noted previously for Band 6 (1.26\,mm) and Band 3 (3\,mm) images; they are also similar to negative \ha\ images of equivalent resolution. Polar coronal holes, which are clearly seen in the 304\,\AA\ band and small \ha\ filaments, are not detectable at 0.86\,mm. We computed the center-to-limb variation (CLV) of the brightness temperature, $T_b$, in Band 7, as well as in Bands 6 and 3, which were obtained during the same campaign, and we combined them to a unique curve of $T_b(\log\mu_{100})$, where $\mu_{100}$ is the cosine of the heliocentric angle reduced to 100\,GHz. Assuming that the absolute calibration of the Band 3 commissioning observations is accurate, we deduced a brightness temperature at the center of the disk of 6085\,K for Band 7, instead of the value of 5500\,K, extrapolated from the recommended values for Bands 3 and 6. More importantly, the $T_b(\log\mu_{100})$ curve flattens at large values of $\mu_{100}$, and so does the corresponding $T_e(\log\tau_{100})$ at large $\tau_{100}$. This is probably an indication that we are approaching the temperature minimum.
}


   \maketitle
%

\section{Introduction}\label{intro}

Our knowledge of the physical conditions of the solar chromosphere is 
primarily based on optical and extreme ultraviolet (EUV) observations.
However, their interpretation is not straightforward because of the 
complexity of these diagnostics. The chromosphere emits at millimeter 
wavelengths as well. Such observations provide simpler
diagnostics because the quiet Sun emission originates from thermal bremsstrahlung
in local thermodynamic equilibrium \citep[LTE; e.g., see][for details]{2016SSRv..200....1W}. However, older millimeter-wavelength data \citep[see 
references compiled by][]{2004A&A...419..747L} suffered from
low spatial resolution and absolute calibration problems, which limited their 
usefulness in modeling. 

\medskip
\medskip
The Atacama Large Millimeter/submillimeter Array (ALMA) has opened a new, 
hitherto underexplored, window for solar observations 
\citep{2017SoPh..292...88W, Shimojo et al. 2017}. Several works using Band 3 
(100\,GHz, 3\,mm) (e.g., \citeauthor{Shimojo et al. 2017}\citeyear{Shimojo et al. 2017}; \citeyear{2020ApJ...888L..28S}; \citeauthor{2017ApJ...845L..19B} \citeyear{2017ApJ...845L..19B}; \citeauthor{2018ApJ...863...96Y} \citeyear{2018ApJ...863...96Y}; \citeauthor{2018A&A...619L...6N} \citeyear{2018A&A...619L...6N}; \citeyear{2020A&A...638A..62N}; \citeauthor{2019A&A...622A.150J} \citeyear{2019A&A...622A.150J}; \citeauthor{2019ApJ...881...99M} \citeyear{2019ApJ...881...99M}; \citeauthor{2019ApJ...877L..26L} \citeyear{2019ApJ...877L..26L}; \citeauthor{2020A&A...635A..71W} \citeyear{2020A&A...635A..71W}; \citeauthor{2020A&A...634A..86P} \citeyear{2020A&A...634A..86P}; \citeauthor{2022ApJ...924..100C} \citeyear{2022ApJ...924..100C}) and Band 6 (239\,GHz, 1.25\,mm) observations \citep[e.g.,][]{2021ApJ...906...82C, 2021A&A...652A..92N, 2021ApJ...920..125M, 2021RSPTA.37900174J} 
have provided new information 
about the upper solar chromosphere, on diverse topics such as the structure 
of the quiet Sun and the chromospheric network, spicules beyond the limb, 
comparison of observations with models of the low chromosphere and with 
radiative magnetohydrodynamic models, weak transient phenomena and oscillations, active region 
plages, and sunspots. In these studies, the high resolution 
interferometric images have been combined with low-resolution full-disk 
images.  We note that higher frequency ALMA bands were not available for solar observations before observing cycle 7 which started in October 2019, although a Band 9 (0.45\,mm) test image, which has not been released yet for science, was presented in \cite{2018Msngr.171...25B}. 

\medskip
In previous works with full-disk images \citep[][hereafter Paper I and Paper II, respectively]{2017A&A...605A..78A, 2020A&A...640A..57A}, we studied the structure of the quiet Sun in ALMA Bands 3 and 6. We reported that the chromospheric network is well visible in ALMA full-disk images, which are similar in structure to Atmospheric Imaging Assembly (AIA) images at 1600\,\AA\ and 304\,\AA\ images. Furthermore, we used the center-to-limb variation of the brightness temperature, $T_b$, to compute the electron temperature, $T_e$, as a function of optical depth at 100\,GHz, $\tau_{100}$, and we compared it with standard atmospheric models.   

\medskip
In this letter we present the first full-disk ALMA observations of the Sun in ALMA Band 7 (347\,GHz, 0.86\,mm). We present our observations in Sect.~\ref{Obs}, our results in Sect.~\ref{Results}, and our conclusions in Sect.~\ref{Summary}. 

\begin{table*}
\begin{center}
\caption{ALMA TP observations from January 2020}
\label{Table:Obs}
\begin{tabular}{lcccl}
\hline 
Date          & Band & Spectral windows & Antennas & Time \\
              &      &  (GHz)           &                 &  (UT) \\
\hline
January 4    & 3   & 93, 95, 105, 107 & 3, 4 &18:07:28, 18:17:47, 18:26:31 \\
            &    &     & &18:35:24, 18:44:16, 18:53:08 \\
            &    &     & &19:01:58, 19:10:50, 19:19:40\\
            &    &     & &19:28:30, 19:37:20\\
\hline
January 4   & 6 & 230, 232, 246, 248 & 3, 4 &12:47:51, 13:01:17, 13:14:41\\
            &   &      & &13:28:06, 13:41:35, 13:55:03\\
            &   &     & &14:08:30, 14:21:59, 14:35:28\\
\hline
January 8   & 7 & 341, 343, 353, 355 & 2 & 14:24:34, 14:40:26, 14:59:46\\
\hline 
\end{tabular}
\end{center}
\end{table*}

\begin{figure*}
\centering
\includegraphics[width=\textwidth]{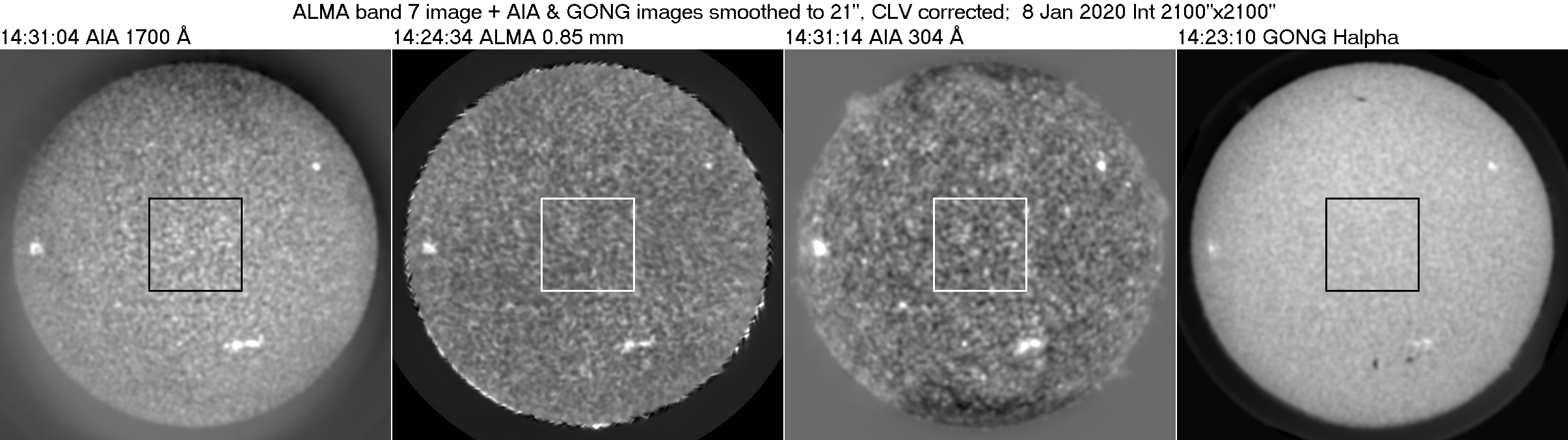}
\caption{Full-disk ALMA Band 7 image at 14:24:34 UT, together with AIA images at 1700\,\AA\ and 304\,\AA, and a GONG \ha\ image. Spoke-wheel structures beyond the limb in the Band 7 image are artifacts due to the scanning process.  All images have been partially corrected for center-to-limb variation (CLV), by subtracting 85\% of the azimuthally averaged intensity. The squares mark the regions shown in Fig.~\ref{partial}; non-ALMA images have been smoothed to the ALMA resolution of 21\arcsec.}
\label{FD}
\end{figure*}

\bigskip
\section{Observations and data reduction}\label{Obs}

In this Letter we analyze full-disk images obtained during an observing 
campaign in which we performed interferometric observations at ALMA's Bands 3, 
6, and 7 (0.86 mm; 347 GHz) of seven quiet 
Sun regions from the center of the disk to the limb. The interferometric 
observations, which will be presented elsewhere,  were carried out in ALMA's 
C-2 configuration which is the one that provides the highest spatial resolution
($\sim$0.7$''$) for Band 7 solar observing programs. The full-disk images 
were obtained by 12-m  ``total power'' (TP) antennas.
Each band was set up in four spectral windows, each having a total bandwidth of
about 2 GHz. More details about the TP observations are presented in Table 1 
which gives the date and time of observations, the central frequency of each 
spectral window, and the labels of the 12-m antennas that acquired the images. 
At any given time, each antenna acquired one image per spectral window; 
therefore, we obtained 44, 36, and 12 images per antenna in Bands 3, 6, and 7,
respectively. In Bands 3 and 6, we only used the images that were provided
by antennas PM03 and PM04, respectively, because of their better system
temperature, $T_{sys}$. The full width at half maximum (FWHM) of the Band 7
single dish beam was 21$''$ and each TP antenna scanned the full solar disk
in 3 min.

\begin{figure*}
\centering
\includegraphics[width=\textwidth]{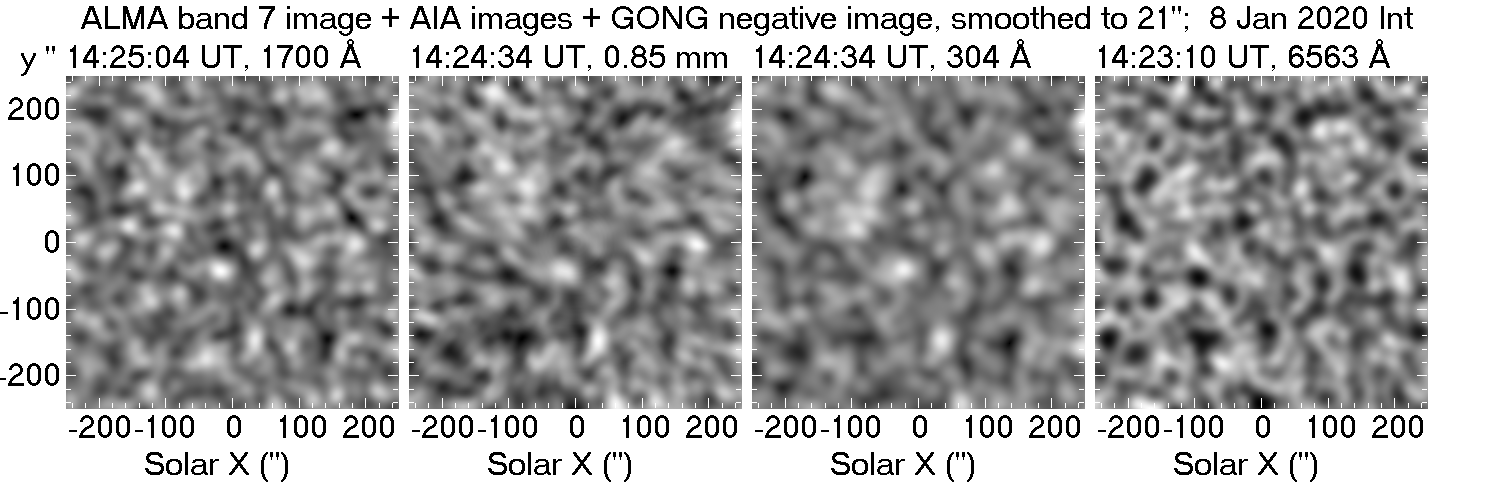}
\caption{Enlarged central region of the images shown in Fig~\ref{FD}; here the negative of the \ha\ image is given.}
\label{partial}
\end{figure*}

The TP data were acquired and processed using the methodology described 
by \cite{2017SoPh..292...88W}. However, the implementation of the Common Astronomy 
Software Applications (CASA) processing scripts provided by the Joint ALMA 
Observatory (JAO) resulted in Band 7 images which suffered from various degrees
of large-scale arc-like brightness modulations which were reminiscent of 
the ``double-circle'' scanning pattern which is used for mapping. It is possible
that these modulations are associated with opacity variations during 
observations. We were able to largely suppress them by invoking CASA's task
sdgaincal before mapping. This task takes advantage of the fact that the 
double-circle mode observes the same position in the center of the field of 
view and adjusts the gains throughout the whole dataset relative to the
measured intensity at the center.

\section{Results and discussion}\label{Results}

\subsection{Structures on the solar disk}\label{fulldisk}
Comparisons of structures in ALMA Bands 3 and 6 with those in other wavelengths have been presented in the past \citep[Paper I;][]{2018A&A...613A..17B} and hence, in this section, we focus on Band 7. A full-disk Band 7 image is shown in Fig.~\ref{FD}; AIA images in the 1700\,\AA\ and 304\,\AA\ bands, as well as a GONG \ha\ image are also displayed for comparison. We note that no full disk AIA { images and no HMI data were available at the time of our ALMA observations}, and thus the nearest ones to the ALMA image are given in the figure; moreover, the 1600\,\AA\ AIA image was affected by flat-field and diffuse light problems, hence the less affected 1700\,\AA\ image is shown instead.

The sun was very quiet on the day of our observations, with a small, spotless active region (NOAA 12755) in the south and two small plage regions in the east and northwest. As expected, these appear brighter than their surroundings in Band 7 by about 900\,K. Two large coronal holes are visible near the poles in the 304\,\AA\ image, with no counterpart in the ALMA image. Three small filaments -- one in the northern and two in the southern hemisphere -- are visible as dark patches near the central meridian in \ha, without any counterpart in the other images. We note that filament channels were quite prominent in Band 3 and Band 6 images obtained in December 2015 \citep[see Paper I, also][]{2018A&A...613A..17B}; moreover, \cite{1993ApJ...418..510B} analyzed 0.85 and 1.25\,mm observations of filaments and prominences and concluded, among other things, that \ha\ filaments are quite optically thin at 0.85\,mm.

Enlarged images of a 500\arcsec\ by 500\arcsec\ region at the center of the disk are shown in Fig.~\ref{partial}, where the negative of the \ha\ image is given. AIA images of this central region were available, and these are shown here instead of the ones in Fig.~\ref{FD}. We note that there is practically a one-to-one correspondence between the ALMA and the AIA network structures, as  previously reported for Bands 6 and 3 by \cite{2017A&A...605A..78A}. Moreover, there is a strong similarity of the ALMA image to the negative \ha\ images; this was first reported by \cite{2021A&A...652A..92N} and attributed to spicules forming above the network elements and seen in absorption in \ha. A detailed discussion of the association of ALMA and \ha\ features \citep[e.g.,][]{{2017A&A...597A.138R, 2017A&A...598A..89R, 2021arXiv210302369R}} cannot be carried out with images at the present resolution; not only does it require high resolution ALMA data, but \ha\ observations of a similar resolution as well \citep[such as those of the plage region imaged by][]{2019ApJ...881...99M}.

\subsection{Center-to-limb variation and modeling}
In previous works (Papers I and II), we used the center-to-limb brightness variation in ALMA Bands 3 and 6, together with older data from \cite{1993ApJ...415..364B} at 353\,GHz, to compute empirical models of the variation of the electron temperature as a function of optical depth, $\tau$. With the present Band 7 observations, we have a homogeneous set of data at our disposal, which we subsequently exploit in this section in order to obtain more reliable results. We note that this approach has a number of advantages over the comparison of the CLV observed at various frequencies to curves obtained from models \citep{2019ApJ...871...45S}, or the determination of limb brightening coefficients \citep{2019SoPh..294..163S}: (a) data at different frequencies are treated simultaneously; (b) there is no need to go close to the limb, thus avoiding primary beam effects; and (c) it provides the physically meaningful quantity $T_e(\tau)$ directly.

The measurements were corrected for diffuse light as described in Paper I.  We avoided measurements close to the limb, going up to 70, 35, and 30\arcsec\ from the limb in Bands 3, 6, and 7, respectively,  which is more than the corresponding beam sizes. Data at different frequencies were combined by reducing all measurements to a common reference frequency, $f_{ref}$,  taking advantage of the fact that both the free-free \citep{1970resp.book.....Z} and the H$^-$ \citep{1974ApJ...187..179S} absorption coefficients are proportional to $f^{-2}$. Hence a measurement of $T_b$ at a frequency $f$ is remapped to
\be
T_b\left((f/f_{ref})^2\mu,f_{ref}\right)=T_b(\mu,f)
\label{reduce}
.\ee
where $\mu=\cos\theta$, with $\theta$ being the heliocentric angle. We note that although the contribution of H$^-$ in the opacity is small ($\sim10$\% around $T_e=6000$\,K), it is not negligible.

\begin{figure}[t]
\centering
\includegraphics[width=\hsize]{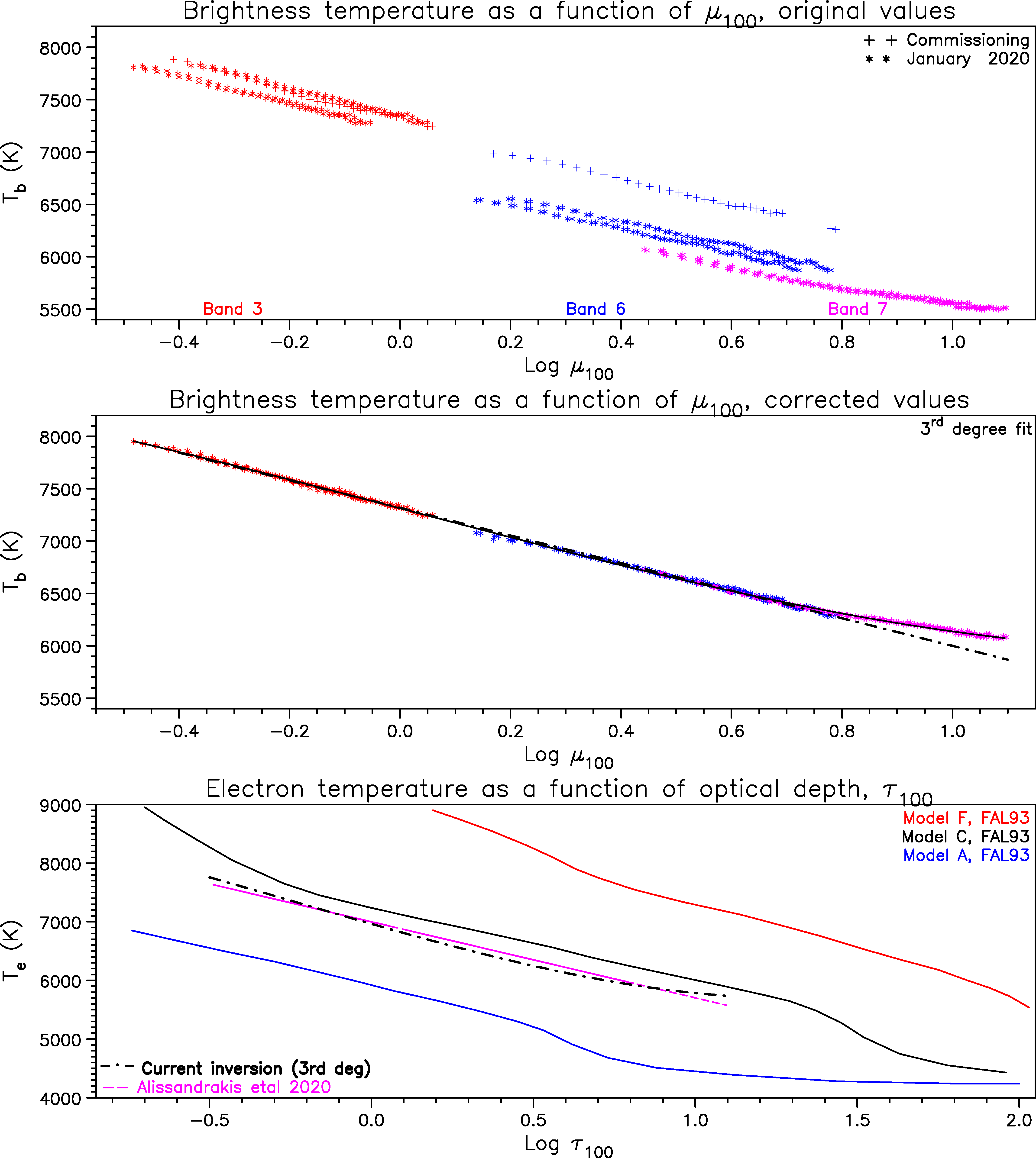}
\caption{CLV observations and their inversion. {\it Top row:} Measured brightness temperature as a function of reference $\mu$, for commissioning data and from the current data set. {\it Middle row:} Normalized data set. The solid line shows a third degree fit and the dash-dotted line shows a linear fit up to $\log\mu_{100}=0.8$.  {\it Bottom row:} Electron temperature as a function of the reference optical depth, deduced from the inversion of the observations. Model curves from \cite{1993ApJ...406..319F} and our results from Paper II are also plotted.}
\label{CLV}
\end{figure}

The results are shown in the top panel of Fig.~\ref{CLV}, where we have included, for comparison, data from the commissioning observations of December 2015, which were also used in Papers I and II. As noted in Paper II, there is a jump in $T_b$ between the commissioning and the current Band 6 data set, attributed to absolute calibration issues, and so is the case with Band 7. Moreover, the Band 7 $T_b(\mu)$ data show a smaller slope compared to Bands 3 and 6, indicating a flattening of the overall  $T_b(\mu)$ curve at higher frequencies.

Working as in Paper II, we normalized Band 6 and 7 measurements through a least-squares fit of all sets to the same polynomial function of $T_b(\log \mu)$. Considering that Band 3 is more reliable \citep{2017SoPh..292...88W}, we set the normalization factor of the commissioning data for that at unity, leaving the factors for the other data sets to be determined by the fit. In order to accommodate for the flattening of the curve in Band 7, we used the following third degree fit:
\be
T_b(\mu)=A_0+A_1\ln\mu+A_2\ln^2\mu+A_3\ln^3\mu \label{logmu0},
\ee
rather than the linear fit used in previous works.

The fit is very good, as shown in the middle panel of Fig.~\ref{CLV}, with the third degree curve standing clearly above the dash-dotted line, which is the extrapolation of the linear fit up to $\log\mu_{100}=0.8$. The root mean square (rms) deviation was about 13\,K, whereas linear and quadratic fits gave rms deviations of 26.6\,K and 16.0\,K, respectively. 

The derived brightness temperatures of the center of the disk are given in Table~\ref{TbCen}, together with the recommended values and those of Paper II, for the four spectral windows of each ALMA band and for the average frequency, together with the percentage differences from the quiet Sun central disk brightness temperatures recommended by \cite{2017SoPh..292...88W} (in bold); as  there is no recommended value for Band 7, the one in the table is the extrapolation of the values recommended for Bands 3 and 6. Also estimates for Band 5 are given in
the table; this band is available for solar work, although no observations have been obtained yet.  The values for Bands 3 and 6 are very close to those given in Paper II, while the disk temperature for Band 7 is 6085\,K, or 10.6\% above the extrapolated recommended value of 5500\,K. 

\begin{table}
\begin{center}
\caption{Disk center brightness temperatures}
\label{TbCen}
\begin{tabular}{lcccccc}
\hline 
Band&Freq&$T_b$, recom\footnotemark&\multicolumn{2}{c}{$T_b$, Paper II}&\multicolumn{2}{c}{$T_b$, this work} \\
&GHz&K&K&\%&K&\%\\
\hline 
Band 3 SW1 & ~93 && 7406&& 7431 \\
Band 3 SW2 & ~95 && 7382&& 7406 \\
\bf{Band 3 Aver}& \bf{100} &\bf{7300}& \bf{7324}&{\bf0.3}& \bf{7347}&{\bf~0.6} \\
Band 3 SW3 & 105 && 7269&& 7289 \\
Band 3 SW4 & 107 && 7248&& 7266 \\
\hline 
Band 5 SW1 & 191 && -- && 6571 \\
Band 5 SW2 & 193 && -- && 6560 \\
\bf{Band 5 Aver}& \bf{198} &\bf{--}& \bf{--}&& \bf{6532} \\
Band 5 SW3 & 203 && -- && 6506 \\
Band 5 SW4 & 205 && -- && 6495 \\
\hline 
Band 6 SW1 & 230 && 6386&& 6382 \\
Band 6 SW2 & 232 && 6376&& 6374 \\
\bf{Band 6 Aver}& \bf{239} &\bf{5900}& \bf{6343}&{\bf7.5}& \bf{6347}&{\bf~7.6} \\
Band 6 SW3 & 246 && 6310&& 6322 \\
Band 6 SW4 & 248 && 6301&& 6315 \\
\hline 
Band 7 SW1 & 341 && --  && 6095 \\
Band 7 SW2 & 343 && --  && 6092 \\
\bf{Band 7 Aver}& \bf{347} & {\bf{5500}} & \bf{--}&& \bf{6085}&{\bf10.6} \\
Band 7 SW3 & 351 && --  && 6077 \\
Band 7 SW4 & 353 && --  && 6074 \\
\hline
\end{tabular}
\end{center}
\footnotesize{$^1$ From \cite{2017SoPh..292...88W}} 
\end{table}

The assumed form of $T_b(\mu)$ of Eq.~(\ref{logmu0}) corresponds to a similar variation of electron temperature with optical depth (see Appendix~\ref{appendix}):
\be
T_e(\tau)=a_0+a_1\ln\tau+a_2\ln^2\tau+a_3\ln^3\tau, \label{logform0}
\ee
the coefficients of which can be evaluated from Eq.~(\ref{invert}) and are given in Table~\ref{parameters}, together with the values from Paper II.

\begin{table}[h]
\begin{center}
\caption{Atmospheric parameters from ALMA inversion}
\label{parameters}
\begin{tabular}{ccc}
\hline 
Parameter&Paper II&This work\\
\hline 
$a_0$  &  6999   & 6964 \\
$a_1$  & $-563$ & $-680$ \\
$a_2$  & -- &  ~~~~30 \\
$a_3$  & -- &  ~~~~19\\
\hline 
\end{tabular}
\end{center}
\end{table}

The resulting $T_e(\tau)$ is plotted in the bottom panel of Fig.~\ref{CLV}, together with the corresponding curves from models A (cell interior), C (average quiet Sun), and F (network) of \cite{1993ApJ...406..319F} and our results from Paper II. Our present curve is very close to that of Paper II, with an important qualitative difference: the flattening at high frequencies. Apparently, this is because in the higher optical depth reached in Band 7, we approach the temperature minimum.

\section{Summary and conclusions}\label{Summary}
In this work, we have presented and studied the first images from ALMA solar sub-millimetric full-disk observations, obtained during a quiet day. The network structures are very similar to those in 1700 and 304\,\AA, as well as in negative \ha\ images, smoothed to the ALMA resolution. Small plage regions are about 900\,K brighter than their surroundings, while the polar coronal holes and small \ha\ filaments are not detectable.

Following the methodology that we developed in Papers I and II, we combined the CLV measurements from Bands 3, 6, and 7 and computed the electron temperature as a function of the optical depth at 100\,GHz, $\tau_{100}$. The curve shows a well-defined flattening at high $\tau_{100}$, which reflects the fact that the CLV curve for Band 7 is flatter than those for Bands 3 and 6. Therefore, a third degree fit was performed for the $T_e(\log\tau_{100})$ curve, rather than the linear fit used in previous works. 

The present result is more reliable since we have a homogeneous data set and thus there was no need to resort to the old measurements of \cite{1993ApJ...415..364B} at 350\,GHz, as we had done before. It is quite plausible that this flattening is due to the fact that in Band 7, we approach the region of the temperature minimum. Still, this result relies on observations during a single day and needs verification with more data. Moreover, the extension of ALMA observations to higher frequencies will provide valuable information on what happens in the deeper chromospheric layers, closer to the temperature minimum. { It is quite unfortunate that the only existing Band 9 full disk image \citep{2018Msngr.171...25B}, which would take us up to $\log\mu_{100}\simeq1.65$, has not been released for science; even so, a visual inspection of that image shows very little, if any, limb brightening, thus affirming our conclusions.} 

In the process of combining the CLV curves, we verified the result of Paper II that the true brightness temperature at the center of the solar disk in Band 6 is above the recommended value; similarly for Band 7, we deduced a disk center brightness of  6085\,K, instead of the extrapolated recommended value of 5500\,K. Reliable as this result may be, it cannot replace the imperative need for absolute calibration of ALMA solar observations, using the moon for example. It is also desirable to have solar observations in ALMA Bands 4 or 5 in order to bridge the gap between Bands 3 and 6.  In anticipation of Band 5 solar observations, we provided estimates of $T_b$ at the center of the disk for the spectral windows of that band as well. We expect additional interesting results from the analysis of the interferometric observations obtained during the same campaign, which is in progress.
   
\begin{acknowledgements}
This paper makes use of the following ALMA data: ADS/JAO.ALMA\#2019.1.01532.S. ALMA is a partnership of ESO (representing its member states), NSF (USA), and NINS (Japan), together with NRC (Canada) and NSC and ASIAA (Taiwan), and KASI (Republic of Korea), in cooperation with the Republic of Chile. The Joint ALMA Observatory is operated by ESO, AUI/NRAO, and NAOJ. The authors are grateful to the AIA and GONG teams for the operation of these instruments and for making available the data to the community.
\end{acknowledgements}

\begin{appendix}

\section{Inversion of the transfer equation for $T_e=f(\ln\tau)$}\label{appendix}
Assuming that

\be
T_e(\tau)=a_0+a_1\ln\tau+a_2\ln^2\tau+a_3\ln^3\tau \label{logform}
,\ee
and substituting into the formal solution of the transfer equation,
\be
T_b(\mu)=\int_0^\infty T_e(\tau)\, e^{-\tau/\mu}\,d\tau/\mu \label{transf}
,\ee
we obtain the expression:
\bea
T_b(\mu)&=&a_0\int_0^\infty e^{-\tau/\mu}\,d\tau/\mu  +a_1\int_0^\infty \ln\tau\,e^{-\tau/\mu}\,d\tau/\mu\\ \nonumber         &+&a_2\int_0^\infty\ln^2\tau\, e^{-\tau/\mu}\,d\tau/\mu+a_3\int_0^\infty\ln^3\tau\, e^{-\tau/\mu}\,d\tau/\mu.
\eea
After the substitution:
\bea
\ln(\tau)&=&\ln(\tau/\mu)+\ln\mu, \hspace{0.3cm}\mbox{ and} \\
x&=&\tau/\mu
\eea 
we have, further,
\bea
T_b(\mu)&=&a_0\int_0^\infty e^{-x}\,dx  +a_1\int_0^\infty (\ln x+\ln\mu)\,e^{-x}\,dx\\ \nonumber         
&+&a_2\int_0^\infty(\ln x+\ln\mu)^2\, e^{-x}\,dx \\ \nonumber
&+& a_3\int_0^\infty(\ln x+\ln\mu)^3\, e^{-x}\,dx
\eea
and, after binomial expansion,
\bea\label{A7}
T_b(\mu)&=&a_0\int_0^\infty e^{-x}\,dx  \\ \nonumber
&+&a_1\ln\mu \int_0^\infty e^{-x}\,dx +a_1\int_0^\infty \ln x\,e^{-x}\,dx \\ \nonumber
&+&a_2\ln^2\mu\int_0^\infty\, e^{-x}\,dx + 2a_2\ln\mu\int_0^\infty\ln x\, e^{-x}\,dx \\ \nonumber
&+&a_2\int_0^\infty\ln^2 x\, e^{-x}\,dx + a_3\ln^3\mu\int_0^\infty\, e^{-x}\,dx\\ \nonumber
&+&3 a_3\ln^2\mu\int_0^\infty \ln x\, e^{-x}\,dx +3 a_3\ln\mu\int_0^\infty \ln^2x\, e^{-x}\,dx \\ \nonumber
&+& a_3\int_0^\infty\ln^3 x\, e^{-x}\,dx.
\eea
The evaluation of the integrals gives:
\bea
\int_0^\infty e^{-x}\,dx &=&1\\
\int_0^\infty \ln x\,e^{-x}\,dx&=&-\gamma=-0.5772157=C_1\\
\int_0^\infty \ln^2 x\,e^{-x}\,dx&=& \gamma^2+\frac{\pi^2}{6}=1.978112=C_2\\
\int_0^\infty \ln^3 x\,e^{-x}\,dx&=&-\left(\gamma^3 +\gamma\frac{\pi^2}{2} + 2\zeta(3)\right)\\ \nonumber
                                                &=&-5.444874=C_3
\eea
where $\gamma$ is the Euler constant and $\zeta$ is the Riemann zeta function.\\
Substituting in (\ref{A7}) and rearranging terms, we obtain:
\bea\label{A12}
T_b(\mu)&=&a_0 + a_1\,C_1 + a_2\,C_2 + a_3\,C_3 \\ \nonumber
&+& (a_1 + 2a_2C_1 + 3a_3C_2)\ln\mu \\ \nonumber
&+& (a_2 +3 a_3\,C_1)\ln^2\mu\\ \nonumber
&+& a_3\ln^3\mu. 
\eea

\noindent 
Comparing (\ref{A12}) with the LSQ fit of the observed $T_b(\mu)$:
\be
T_b(\mu)=A_0+A_1\ln\mu+A_2\ln^2\mu+A_3\ln^3\mu \label{logmu},
\ee
we obtain the coefficients of $T_e(\tau)$:
\bea
\label{invert}
a_3&=&A_3\\ \nonumber 
a_2&=&A_2 - 3 a_3\,C_1 \\ \nonumber
a_1&=&A_1 - 2a_2C_1 - 3a_3C_2 \\ \nonumber
a_0&=&A_0 - a_1\,C_1 - a_2\,C_2 - a_3\,C_3. 
\eea
\end{appendix}

\listofobjects

\end{document}